\documentclass[aip,pop,reprint]{revtex4-1}
\usepackage{amsmath}
\usepackage{amssymb}
\usepackage{color}
\usepackage{epsfig}
\usepackage{graphicx}
\usepackage{bm}
\usepackage{natbib}
\usepackage{algorithmic}
\usepackage{algorithm}
\usepackage{url}

\begin{document}

\title{The Generation of Random Variates From a Relativistic Maxwellian
Distribution}

\author{M. Swisdak}
\email{swisdak@umd.edu}
\affiliation{Institute for Research in Electronics and Applied Physics,
University of Maryland, College Park, MD 20742}

\date{\today}

\begin{abstract}
A procedure for generating random variates from a relativistic
Maxwellian distribution with arbitrary temperature and drift velocity
is presented.  The algorithm is based on the rejection method and
can be used to initialize particle velocities in kinetic simulations
of plasmas and gases.
\end{abstract}


\maketitle

\section{Introduction}\label{intro}

Kinetic simulations of plasmas and gases follow the trajectories of
individual particles as they evolve from an initial state, often one
for which the velocities are random variates drawn from a specified
distribution function.  For systems with negligible relativistic and
quantum effects, that distribution is often the Maxwell-Boltzmann
(Maxwellian).  Generating initial particle velocities is then
straightforward since a non-relativistic Maxwellian separates, with
each coordinate governed by a Gaussian (normal) distribution for which
quick and simple methods of generating variates are known
\cite{marsaglia64a}.  Drifting distributions (those with non-zero
average velocities) are only slightly more complicated since Galilean
transformations apply in the non-relativistic limit.

Special relativity, however, does not permit velocities greater than
the speed of light $c$ and is thus incompatible with a Maxwellian
distribution that predicts a non-zero probability for every velocity.
J\"uttner \cite{juttner11a} introduced the first relativistic
generalization of the Maxwellian, basing the derivation on Boltzmann's
expression for the probability of finding a system in a state with
energy $E$,
\begin{equation}
f  = \frac{1}{Z(T)} e^{-E/k_BT}
\end{equation}
Here $f$, $k_B$, $T$, and $Z(T)$ are the distribution function,
Boltzmann's constant, the temperature, and the partition function,
respectively.  In special relativity the total energy for a particle
with rest mass $m$ is $E =
\gamma mc^2$, although it is often more useful to consider the kinetic
energy $(\gamma-1)mc^2$, which only differs by a factor that can be
absorbed into the normalization.  Defining
\begin{equation}\label{defA}
A = \frac{mc^2}{k_BT}
\end{equation}
and henceforth writing velocities, momenta, and energies in
dimensionless units --- $u = u/c$, $p = p/mc$, and $E = E/mc^2$ --- 
gives the distribution
\begin{equation}\label{boltz}
f(\mathbf{p})  = \frac{1}{Z(T)} e^{-A(E(\mathbf{p})-1)}
\end{equation}
Perhaps surprisingly, the correct form of $Z$ is still being debated
\cite{lehmann06a,chaconacosta10a,treumann11a},
but, fortunately for the purposes of variate generation, it only
contributes a multiplicative factor that, crucially, does not depend
on the momentum.  Hence, for any variate-generating algorithm in which
only ratios of $f$ appear, $Z$ (and, in fact, any multiplicative
pre-factors that do not depend on the momentum) can be dropped.

Neither J\"uttner's distribution nor equation \ref{boltz}, from which
it can be derived, are Lorentz-covariant (i.e., expressed solely in
terms of scalars, 4-vectors, etc.).  Synge \cite{synge57a} was perhaps
the first to suggest a covariant form, the only features of which that
matter for variate generation are: (1) $E$ changes between reference
frames via a Lorentz transformation, and (2) The form of the variation
in $T$ between reference frames is known.  This work assumes $T$ is
an invariant, but the formalism directly carries over to the case
$T=T(u)$, where $u$ is the reference frame velocity.

Both the isotropic and drifting relativistic Maxwellians are special
cases of a class known as generalized hyperbolic distributions that
was first used by Barndorff-Nielsen to describe measurements of
aeolian sand grains \cite{barndorff-nielsen77a}.  Although the
original work considered univariate (i.e., one-dimensional)
distributions, multivariate versions quickly followed
\cite{barndorff-nielsen82a, blaesild81a}, as did algorithms for
generating random hyperbolic variates via a mixture of
normally-distributed variates and variates of the generalized inverse
Gaussian distribution
\cite{atkinson82a}.  The semi-heavy tails of generalized hyperbolic
distributions make them useful for modelers of financial markets and
has spurred the development of several numerical packages for
generating variates.  However they tend to work efficiently only for a
narrow range of temperatures and drift speeds.

Pozdnyakov et al. \cite{pozdnyakov77a,pozdnyakov83a} describe an
algorithm specific to the relativistic Maxwellian distribution, albeit
restricted to the isotropic case, that separately considers the large
and small $A$ limits.  Its efficiency (defined as how often the
procedure generates an acceptable variate), varies from nearly $100\%$
for $A<1$ to $33\%$ for $A\rightarrow \infty$.  The algorithm
presented here is slightly more complicated but also more general, as
it works efficiently ($\approx 90\%$ for all $A$ in the isotropic
case) for nearly arbitrary drift speeds and temperatures

Sections \ref{statmax} and \ref{movemax} address the generation of
variates for stationary and drifting Maxwellians, respectively.  Both
use an algorithm originally due to Devroye \cite{devroye86a} that has
been adapted for this work and is described in the Appendix.

\section{Stationary Maxwellian}\label{statmax}

Using the rest-frame relationship between energy and momentum,
equation \ref{boltz} becomes
\begin{equation}\label{statmech}
f(\mathbf{p})  = e^{-A\left(\sqrt{1+|\mathbf{p}|^2}-1\right)}
\end{equation}
$Z$ has been dropped since, as discussed in Section \ref{intro}, it
does not affect the generation of variates.  This distribution
function is isotropic and can be re-written more conveniently in terms
of $p = |\mathbf{p}|$ by moving to spherical coordinates, integrating
over the angle variables, and dropping the constant pre-factors:
\begin{equation}\label{isomax}
f(p) =  p^2e^{-A\left(\sqrt{1+p^2}-1\right)}
\end{equation}
By keeping the seemingly extraneous $e^A$ term, $f$ can be written in
an equivalent form
\begin{equation}\label{isomax_stable}
f(p) =  p^2e^{-Ap^2/\left(1+\sqrt{1+p^2}\right)}
\end{equation}
that avoids numerical problems in the exponent for small $p$.  Once a
$p$ has been generated from this distribution, it can be uniformly
distributed over the surface of a sphere (see, e.g.,
Knop\cite{knop70a}) to give individual Cartesian components.

\begin{figure}
\includegraphics[width=\columnwidth]{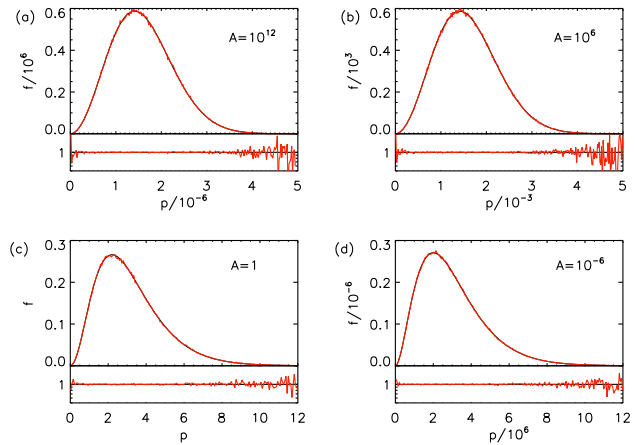}
\caption{\label{isofig} Comparison between the theoretical distribution of
$p$ from equation \ref{isomax_stable} and the generated random
variates for isotropic relativistic Maxwellians with different values
of $A$.  The top plot in each of the four panels shows the theoretical
distribution in black with the computed distribution over-plotted in
red.  The bottom plot shows the ratio of the computed to theoretical
distributions on a log scale ranging from $1/2$ to $2$.  Note the
different axis scales for each panel.}
\end{figure}

Since $f$ is log-concave, i.e., $(\log f)^{\prime\prime} \leq 0$ for
all $p$, the algorithm described in the Appendix can be used to
generate random variates.  The only additional information needed is
the mode
\begin{equation}
p_m^2 = \frac{2}{A^2}\left(1+\sqrt{1+A^2}\right)
\end{equation}
Figure \ref{isofig} demonstrates the results for $A = 10^{12}$
(roughly typical of water vapor at room temperature), $10^6$ (keV
protons), $1$ (MeV electrons), and $10^{-6}$ (TeV electrons).  The top
portion of each panel compares the expected $f$ from equation
\ref{isomax_stable} to a histogram of the generated variates.  
The two curves, each normalized to unit area, are nearly
indistinguishable and so the bottom portion of each panel plots the
ratio of the histogram to the expected value on a logarithmic scale
ranging from $1/2$ to $2$.  Because the variate-generating algorithm
is based on the rejection method, not every generation attempt
succeeds.  For each panel $10^6$ attempts were made, with the success
rate varying only slightly with $A$, from $88\%$ for $A=10^{12}$ to
$90\%$ for $A=10^{-6}$.  Although the algorithm produces excellent
results for all $A$, note that in systems with $A \lesssim 1$
pair production can occur and any distribution that assumes a constant
particle number, such as the $f$ of equation \ref{isomax}, cannot be
completely accurate.

%\begin{center}
%\end{center}

In the non-relativistic limit the exponential term in equation
\ref{isomax} reduces to $\exp(-Ap^2/2)$, which implies that varying
$A$ and $1/p^2$ by the same factor will produce similar distributions.
This effect can be observed in panels (a) and (b) which, aside from
the axis scales, are nearly identical.  Panels (c) and (d) cover the
relativistic limit in which the exponential term reduces to
$\exp(-Ap)$ so that jointly scaling $A$ and $1/p$ produces the same
effect.

\section{Drifting Maxwellian}\label{movemax}

In a frame moving with velocity $\mathbf{u}$ with respect to the rest
frame the Lorentz-transformed energy of a particle is
\begin{equation}
E^{\prime} = \gamma_u(E - \mathbf{u}\boldsymbol{\cdot}\mathbf{p})
\end{equation}
where $\gamma_u = 1/\sqrt{1-u^2}$. Equation \ref{boltz} then becomes
\begin{equation}\label{driftmax}
f(\mathbf{p}) = e^{-A\left(\gamma_u\sqrt{1+|\mathbf{p}|^2} -
\gamma_u\mathbf{u}\boldsymbol{\cdot}\mathbf{p}-1\right)}
\end{equation}
To proceed, define a coordinate system with axes parallel and
perpendicular to the drift velocity, $(p_{\parallel},
\mathbf{p}_{\perp}$).  The anisotropic, non-separable form of $f$
means that the perpendicular momenta must be generated conditional
on the value of $p_{\parallel}$.  Moving to cylindrical coordinates
and integrating over the perpendicular coordinates gives the
distribution for $p_{\parallel}$
\begin{equation}
f(p_{\parallel}) = \int_0^{2\pi}\!\!\!\int_0^{\infty} p_{\perp}
e^{-A\left(\gamma_u\sqrt{1 + p_{\parallel}^2 + |\mathbf{p}_{\perp}|^2}
-
\gamma_uup_{\parallel} - 1\right)}
\,\mathrm{d}p_{\perp}\mathrm{d}\theta
\end{equation}
Ignoring, as usual, multiplicative scaling coefficients leaves
\begin{equation}\label{ppar}
f(p_{\parallel}) = \left(1 +
A\gamma_u\sqrt{1+p_{\parallel}^2}\right)e^{-A \left(
\gamma_u\sqrt{1+p_{\parallel}^2} -\gamma_uup_{\parallel} - 1\right) }
\end{equation}
and it is again useful to express the exponent in a numerically stable
form,
\begin{equation}
f(p_{\parallel}) = \left(1 +A\gamma_u\gamma_{\parallel}\right)
e^{-A(p_{\parallel}-p_u)^2/(\gamma_{\parallel}\gamma_u+p_{\parallel}p_u+1)}
\end{equation}
where $p_u = \gamma_uu$ and $\gamma_{\parallel} =
\sqrt{1+p_{\parallel}^2}$.  This distribution is log-concave with mode
\begin{equation}
p_m = \frac{p_u}{A}\left(1+\sqrt{u^2+A^2}\right)
\end{equation}
and the algorithm of the Appendix will generate variates of
$p_{\parallel}$.

Given a $p_{\parallel}$ variate, a $p_{\perp} = |\mathbf{p_{\perp}}|$
variate conditional on it can be found.  Writing $p_s =
p_{\perp}/\sqrt{1+p_{\parallel}^2}$ the conditional distribution is
\begin{equation}\label{pperp}
f(p_s\,|\,p_{\parallel}) =
p_s\,e^{-A\left(\gamma_u\sqrt{1+p_{\parallel}^2}\sqrt{1+p_s^2} -
\gamma_uup_{\parallel} - 1\right)}
\end{equation}
which is equivalent to
\begin{equation}
f(p_s\,|\,p_{\parallel}) = p_s\,
e^{-A[(p_{\parallel}-p_u)^2+\gamma_{\parallel}^2\gamma_u^2p_s^2]
/(\gamma_{\parallel}\gamma_u\gamma_s+p_{\parallel}p_u+1)}
\end{equation}
The mode of $f$, which is again log-concave, is
\begin{equation}
\qquad p_m^2 =
\frac{1}{2(A\gamma_u\gamma_{\parallel})^2} 
\left(1+\sqrt{1+(2A\gamma_u\gamma_{\parallel})^2}\right)
\end{equation}

With this background, the procedure for generating the variates of a
drifting relativistic Maxwellian is: (1) Generate a variate for
$p_{\parallel}$ using the distribution of equation \ref{ppar}; (2) For
each $p_{\parallel}$ variate, attempt to generate a $p_s$ variate from
the distribution of equation \ref{pperp} (if the attempt fails,
discard $p_{\parallel}$ and return to step 1); (3) With $p_{\perp} =
p_s\sqrt{1+p_{\parallel}^2}$ and a variate $\Theta$ uniformly
distributed on $[0,2\pi]$, generate the individual perpendicular
variates via $p_{\perp}\cos\Theta$ and $p_{\perp}\sin\Theta$; (4) If
necessary, transform from parallel/perpendicular coordinates to the
desired coordinate system.

In order to check this procedure, it is useful to establish a
one-dimensional version of the distribution of equation
\ref{driftmax}.  To do so, express $\mathbf{p}$ in spherical
coordinates, integrate over solid angle, and drop unnecessary scaling
factors:
\begin{equation}
f(p) = p^2e^{-A(\gamma_u\sqrt{1+p^2}-1)}\int_{-1}^{1}
e^{Ap_up\cos\theta}\,\mathrm{d}(\cos\theta)
\end{equation}
The result,
\begin{equation}\label{shiftcomp}
f(p) = p\,\frac{\sinh\left(Ap_up\right)}
{Ap_u}e^{-A(\gamma_u\sqrt{1+p^2}-1)}
\end{equation}
reduces, as it must, to equation \ref{isomax} when $u=0$. 

\begin{figure}
\includegraphics[width=\columnwidth]{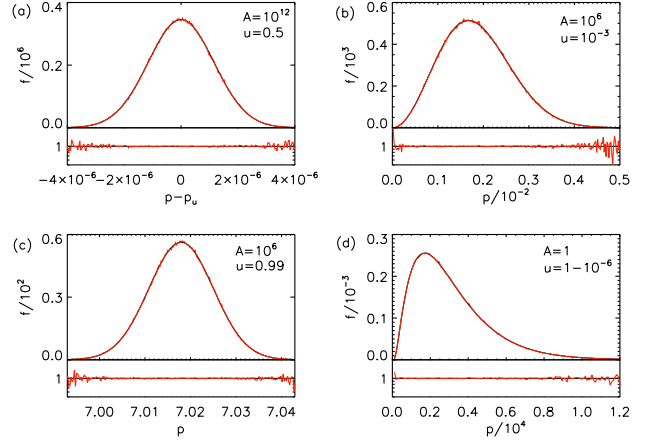}
\caption{\label{shiftfig} Comparison, in the same format as Figure
\ref{isofig}, between the theoretical distribution of $p$ from equation
\ref{shiftcomp} and the generated random variates for drifting
relativistic Maxwellians with different values of $A$ and $u$.  In
panel (a), $p_u = \gamma_uu\approx 0.577350$.  Note the different axis
scales for each panel.}
\end{figure}

Figure \ref{shiftfig} compares the generation of variates from
distributions with four combinations of $A$ and $u$ with the
theoretical form given by equation \ref{shiftcomp}.  For each
combination, $10^6$ attempts were made to generate $p_{\parallel}$
and, if successful, $p_{\perp}$.  The resulting total efficiency of
$77-80\%$ depends only weakly on the parameters with small $A$ (as in
the stationary case) and small $u$ yielding slightly higher values,
reflecting the slightly better fits of the rejection method's bounding
functions in those cases.  In panels (a) and (c) the momentum
corresponding to the drift, $p_u =
\gamma_uu$, is larger than the typical thermal momentum and, as a
consequence, the particle momenta are distributed nearly symmetrically
about $p_u$.  Panels (b) and (d), in contrast, display cases where the
drift has minimal effect and the curves more resemble those of Figure
\ref{isofig}.

%\begin{center}
%\end{center}

The procedure outlined in this section still works when $u=0$, albeit
more slowly and less efficiently than that of Section \ref{statmax}
because it requires a second, nested call of the variate-generating
algorithm.  Interestingly, there exists an analytic expression that,
given a single uniformly distributed variate, always returns a variate
of $p_s = p_{\perp}/\sqrt{1+p_{\parallel}^2}$ from the $f$ of equation
\ref{pperp}.  Such a formula can be found whenever the inverse
function of the integral of $f$ is known; in this case it can be
written in terms of the Lambert $W$ function, for which the defining
equation is $x = W(x)\exp[W(x)]$ (i.e., $W(x)$ is the inverse function
of $y = xe^x$).  The solution is
\begin{equation}\label{lambert}
\gamma_s = -\frac{1}{A\gamma_{\parallel}\gamma_u}\left(1 +
W_{-1}\left[\left(-1-A\gamma_{\parallel}\gamma_u\right)e^{\left(-1 -
A\gamma_{\parallel}\gamma_u\right)}U\right]\right)
\end{equation}
where $U$ is a uniformly distributed variate on $[0,1]$.  $W(x)$ is
multi-valued and $W_{-1}$ refers to the lower branch which, for real
$x$, is only defined on the interval $[-1/e,0)$ with $W_{-1}(-1/e) =
-1$ and $W_{-1}(x\rightarrow 0^-)\rightarrow -\infty$.  Equation
\ref{lambert} can be further simplified by noting that $W_{-1}(x) =
\log(-x) - \log(-W_{-1}(x))$, but even in that form remains ill-suited
for numerical work in the large $A$ limit.

\begin{acknowledgments}
We thank J. Dahlin for helpful discussions and NSF grant PHY1102479.
\end{acknowledgments}

\appendix*
\section{}\label{appendix}

The algorithm below, originally described in Devroye \cite{devroye86a}
and reproduced here in a slightly altered and simpler form, generates
random variates for a log-concave distribution function $f$ (one
satisfying $(\log f)^{\prime\prime}
\leq 0$) by applying the rejection method \cite{press92a} to a
piecewise-continuous upper bound constructed of three functions: a
constant intersecting the mode of $f$ and two exponential tails.
Preliminary calculations are necessary to find the points $p_-$ and
$p_+$ at which the three bounding curves meet; it can be shown that
the optimal choices satisfy $f(p_\pm) = f(p_m)/e$, where $p_m$ is the
distribution's mode.  Every $f$ considered in this work is a
well-behaved uni-modal function and standard root-finding methods
(applied, for computational ease, to the equations
$\log[f(p_\pm)/f(p_m)]+1=0$) quickly converge to $p_\pm$.  Note that
because $f$ only appears in ratios, even during the preliminary
root-finding, its normalization is irrelevant.

Although the algorithm is not guaranteed to generate a variate on each
iteration, its efficiency for an arbitrary $f$ can be shown to lie
between $\approx 30$ and $100\%$.  Generation of $10^6$ variates takes
$\mathcal{O}(0.1)$ seconds in unoptimized Fortran.  For additional
algorithmic details, derivations, a discussion of efficiency, and
alternative methods, see Devroye\cite{devroye86a}.

\newpage

\begin{algorithm}[H]
\caption{Generate a random variate for $f$, a log-concave distribution
function}
\begin{algorithmic}
\REQUIRE $p_m$ is the mode of $f$ \\
$p_+$ and $p_-$ satisfy $f(p_\pm) = f(p_m)/e$ \\ $\lambda_+ \leftarrow
-f(p_+)/f^{\prime}(p_+)$, $\lambda_-
\leftarrow f(p_-)/f^{\prime}(p_-)$ \COMMENT{can be re-written
in terms of $(\log f)^{\prime}$} \\
$q_- \leftarrow \frac{\lambda_-}{p_+-p_-}$, $q_+ \leftarrow
\frac{\lambda_+}{p_+-p_-}$, $q_m \leftarrow 1 - (q_++q_-)$
%\ENSURE $y = x^n$
\REPEAT
\STATE generate $U$ and $V$, uniform variates on $[0,1]$
  \IF{$U \leq q_m$}
  \STATE $Y \leftarrow U/q_m$
  \STATE $X \leftarrow (1-Y)(p_-+\lambda_-) + Y(p_+ - \lambda_+)$
    \IF{$V \leq f(X)/f(p_m)$}
      \STATE done
    \ENDIF
  \ELSIF{$U \leq q_m+q_+$}
  \STATE $E \leftarrow -\log\left(\frac{U-q_m}{q_+}\right)$
  \STATE $X \leftarrow p_+-\lambda_+(1-E)$
    \IF{$V \leq e^Ef(X)/f(p_m)$}
      \STATE done
    \ENDIF
  \ELSE
  \STATE $E \leftarrow -\log\left(\frac{U-(q_m+q_+)}{q_-}\right)$
  \STATE $X \leftarrow p_-+\lambda_-(1-E)$
    \IF{$V \leq e^Ef(X)/f(p_m)$}
      \STATE done
    \ENDIF
   \ENDIF
\UNTIL{done}
\RETURN $X$
\end{algorithmic}
\end{algorithm}


%\bibliography{paper}
%\bibliographystyle{elsarticle-num}

\end{document}